\documentclass[11pt]{article}
\usepackage{moriond}
\usepackage{textcomp}
\usepackage[left]{lineno}
\usepackage{comment}

\def\Journal#1#2#3#4{{#1} {\bf #2}, #3 (#4)}

\def\NIMA{{\em Nucl. Instrum. Methods} A}

\def\PRD{{\em Phys. Rev.} D}

\def\be{\begin{equation}}
\def\ee{\end{equation}}
\def\bea{\begin{eqnarray}}
\def\eea{\end{eqnarray}}

\newcommand{\Photo}{}

\begin{document}
\vspace*{4cm}
\title{Dark Matter directional detection with MIMAC}

\author{Q.~Riffard,  G.~Bosson, O.~Bourrion, O.~Guillaudin, J.~Lamblin, F.~Mayet, J.-F.~Muraz, J.-P.~Richer, D.~Santos}
\address{LPSC, Universit\'e Joseph Fourier Grenoble 1, CNRS/IN2P3, Grenoble INP\\
53 rue des Martyrs, Grenoble, France}
\author{J.~Billard,}
\address{LPSC, Universit\'e Joseph Fourier Grenoble 1, CNRS/IN2P3, Grenoble INP\\
53 rue des Martyrs, Grenoble, France, Department of Physics, Massachusetts Institute of Technology, Cambridge, MA 02139, USA and MIT Kavli Institute for Astrophysics and Space Research, Massachusetts Institute of Technology; Cambridge, MA 02139, USA}
\author{L. Lebreton, D. Maire }
\address{Laboratoire de M\'etrologie et de Dosim\'etrie des Neutrons, IRSN Cadarache, 13115 Saint-Paul-Lez-Durance, France}
\author{J. Busto, J. Brunner, D. Fouchez}
\address{CPPM, Aix-Marseille Universit\'e, CNRS/IN2P3, Marseille, France}

\maketitle

\abstracts{Directional detection is a promising direct Dark Matter (DM) search strategy. The angular distribution of the nuclear recoil tracks from WIMP events should present an anisotropy in galactic coordinates. 
This strategy requires both a measurement of the recoil energy with a threshold of about 5~keV and 3D recoil tracks down to few millimeters. \\
The MIMAC project, based on a \textmu-TPC matrix, with $CF_4$ and $CHF_3$, is being developed. In June 2012, a bi-chamber prototype was installed at the LSM (Laboratoire Souterrain de Modane).
A preliminary analysis of the first four months data taking
 allowed, for the first time, the observation of recoils from the $\mathrm{^{222}Rn}$ progeny.}

\section{Introduction: Directional detection}

Dark Matter (DM) directional detection is based on the idea that the angular distribution of WIMP (Weakly Interacting Massive Particle) momentum directions should present an anisotropy in galactic coordinates~\cite{Spergel:1987kx}. This anisotropy is due to the relative motion of the solar system with respect to the cold galactic DM halo.
Thus, the angular distribution of recoils produced by scattering of WIMPs on nuclei should present an anisotropy
pointing towards the constellation Cygnus. Background events are instead isotropically distributed in Galactic coordinates.
Using a profile likelihood analysis~\cite{billard:2012ProfLikeli} it has been shown that it is possible to extract a DM signal from background events.
As other directional detection experiments~\cite{Ahlen:2009ev}, the aim of the MIMAC project is the measurement of nuclear recoil energy and angular distribution to search for this signature.

\section{MIMAC detection strategy}
\label{sec:Detector}

\begin{figure}[h]
\begin{minipage}{0.65\linewidth}
\centerline{\includegraphics[width=\linewidth]{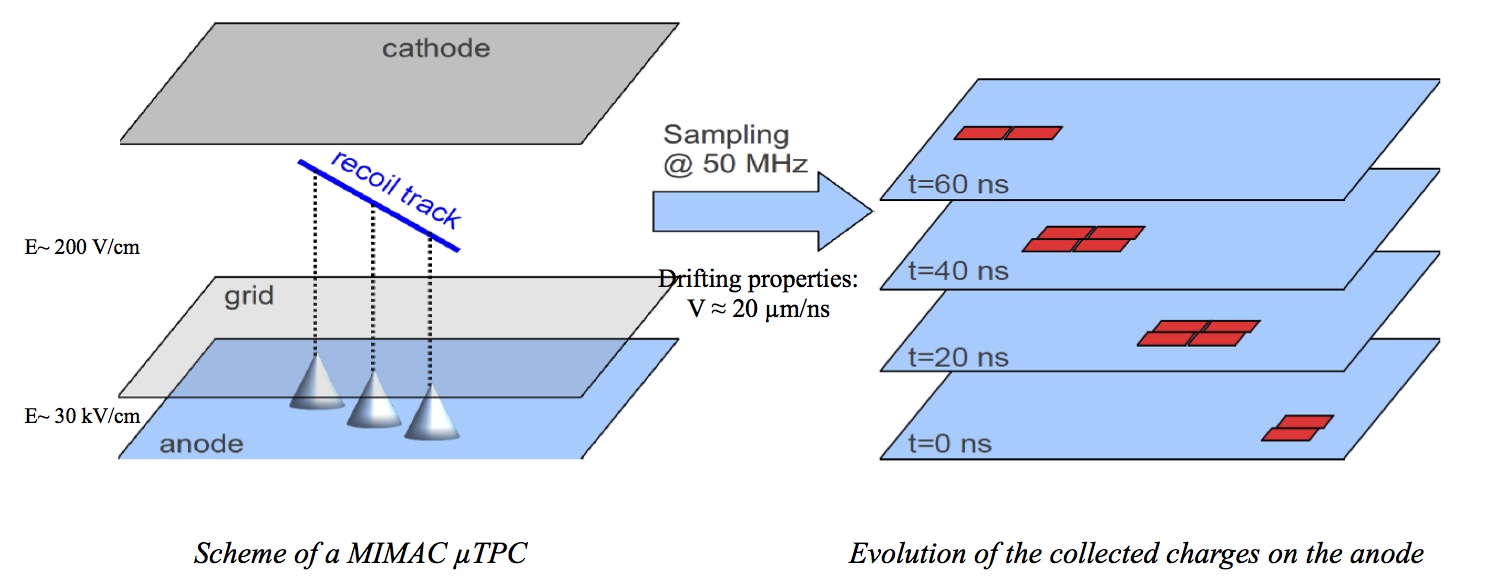}}
\end{minipage}
\hfill
\begin{minipage}{0.33\linewidth}
\centerline{\includegraphics[width=0.9\linewidth]{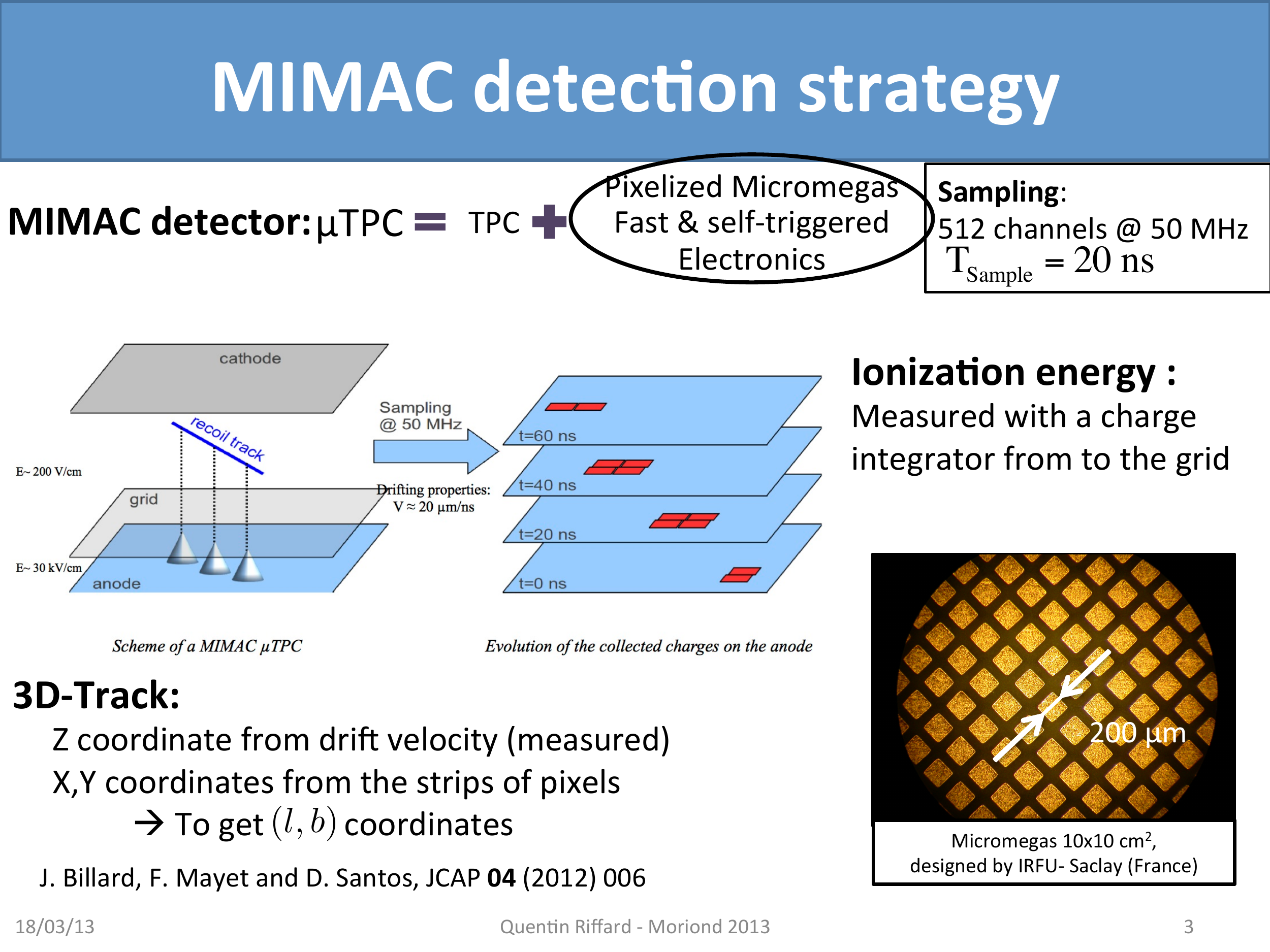}}
\end{minipage}
\caption{Left: Scheme representing a nuclear recoil produced in the active volume of the detector and the sampling of the pixelized anode at $50\, \mathrm{MHz}$ of collected charges. Right: A picture of a small part of the micromegas pixels designed by IRFU ‐ Saclay (France) showing the 200 \textmu m wide pixels.}
\label{fig:ShemaMicromegas}
\end{figure}

The MIMAC detector is a low pressure \textmu-TPC (Time Projection Chamber).
When charged particles pass through the gas detector (e.g. ${\mathrm{CF_4}}$), they lose a part of their energy by ionization, creating pairs of electrons and ions. These electrons are drifted to the grid by an electric field ($E_{\mathrm{drift}} \sim 200\,\mathrm{V.cm^{-1}} $).
Finally, after passing through the grid, these electrons are amplified by avalanche with a high electric field ($E_{\mathrm{aval}} \sim 30\,\mathrm{kV.cm^{-1}} $) reaching the pixelized anode. Fig. ~\ref{fig:ShemaMicromegas} right shows a small part of a $10 \times 10\, \mathrm{cm^2}$ pixelized micromegas~\cite{Iguaz:2011yc} each pixel being 200~\textmu m wide. 
By coupling a micro-pattern detector, such as a micromegas, with a fast and self-triggered electronics~\cite{Bourrion:2011vk}, we are able to sample the $512$ channels of the anode at $50\, \mathrm{MHz}$.
Fig.~\ref{fig:ShemaMicromegas} left illustrates the ionization energy deposition, the charge collection and the sampling of a recoil track every $20\,\mathrm{ns}$. 
The energy released by ionization is measured by a charges integrator (flash-ADC) connected to the grid.
Knowing the drift velocity, nearly 20 \textmu m/ns, which is also measurable with this setup~\cite{Billard:2013cxa}, a 3D track can then be reconstructed~\cite{Billard:2012bk}.

\section{MIMAC bi-chamber prototype at the Laboratoire Souterrain de Modane}

\begin{figure}[h]
\begin{center}
\begin{minipage}{0.4\linewidth}
\centerline{\includegraphics[width=\linewidth]{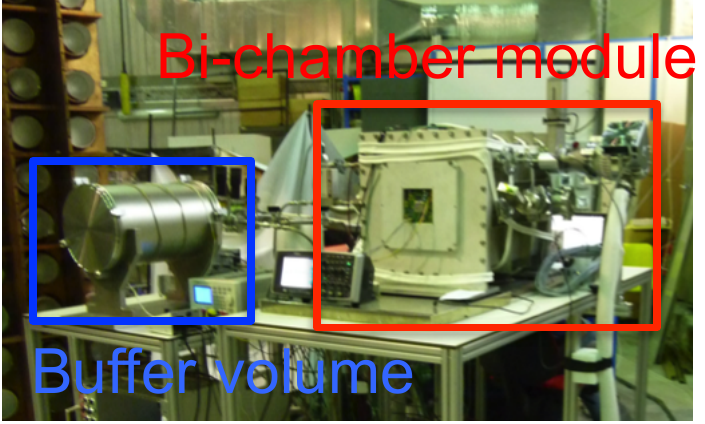}}
\end{minipage}
\begin{minipage}{0.4\linewidth}
\centerline{\includegraphics[width=\linewidth]{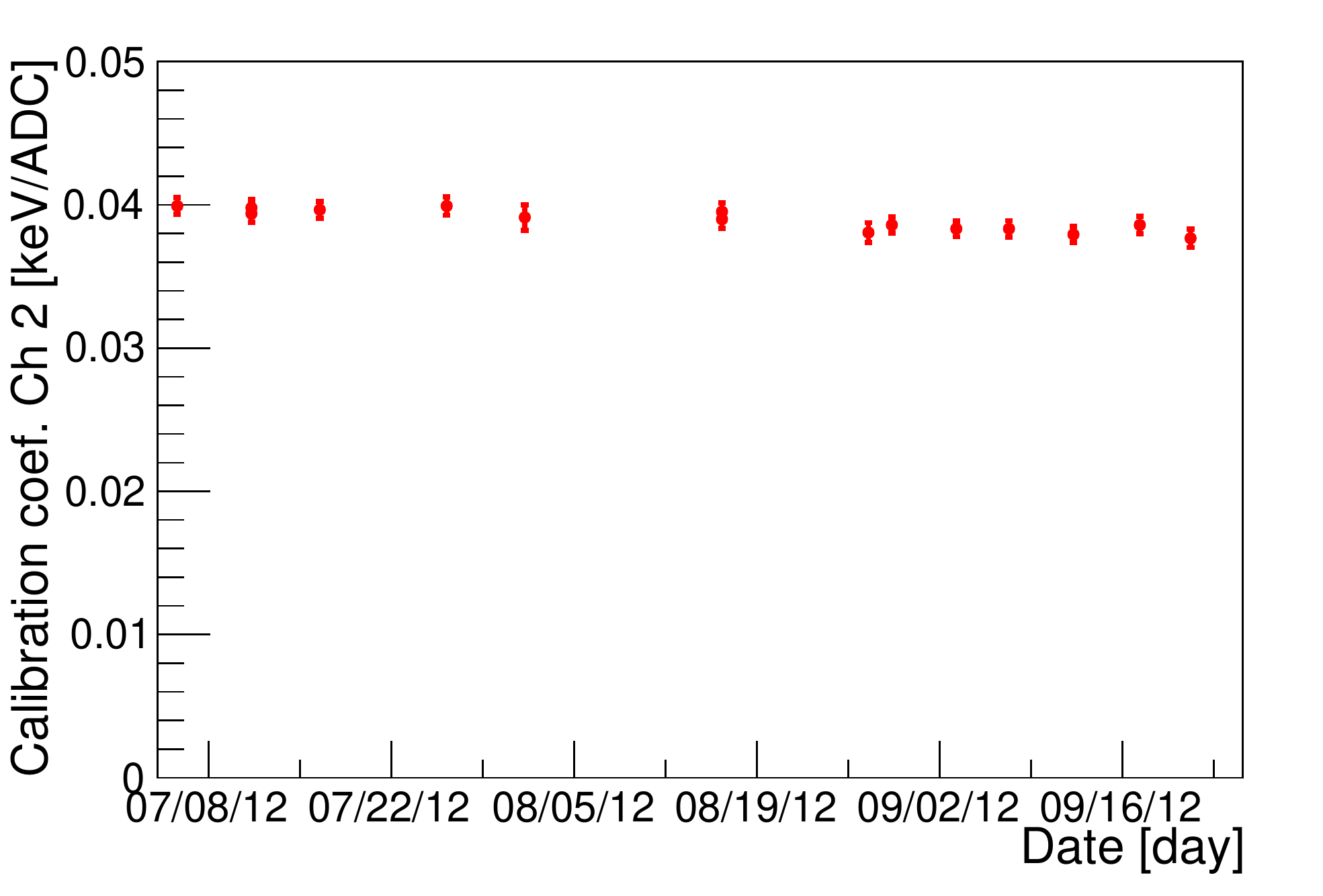}}
\end{minipage}
\caption{Left: The bi-chamber prototype at the Laboratoire Souterrain de Modane in June 2012. The bi-chamber module is identified in red and the buffer volume in blue. Right: The slope coefficient of a linear calibration as a function of time in keV/ADC-Channel. Each measurement has its error bar.}
\label{fig:detect}
\end{center}
\end{figure}

The bi-chamber module shown in Fig.~\ref{fig:detect} left, is composed of two mirrored detectors sharing the same cathode. The prototype active volume ($V  \sim 5.8\,\mathrm{L}$) is filled with the following gas mixture : $\mathrm{ 70\,\%\, CF_4  + 28\,\%\, CHF_3 + 2\,\% \,C_4H_{10}}$  at a pressure of $50\,\mathrm{mbar}$.
The MIMAC bi-chamber prototype was installed at the Laboratoire Souterrain de Modane (LSM) in June 2012 for four months of data taking in an underground laboratory.
By means of an X-ray generator, the detector has been weekly calibrated with fluorescence photons from metal foils.
The gas circulation system includes a buffer volume, an oxygen filter, a dry pump and a pressure regulator. It allows to keep the gas quality stable in a closed circuit. Indeed, the presence of impurities and $\mathrm{O_2}$ must be controlled to
 to prevent gain and resolution degradation of the detector.

Fig.~\ref{fig:detect} right shows the slope coefficient of  a linear calibration as a function of time, highlighting the gain stability during this data taking period.

%%%%%%%%%%%%%%%%%%%%%%%%%%%%%%%%%%%%%%%%%%%%%%%%

\section{Preliminary analysis of the first months of data taking}
\label{sec:Modane}

\begin{figure}[h]
\begin{minipage}{0.5\linewidth}
\centerline{\includegraphics[width=\linewidth]{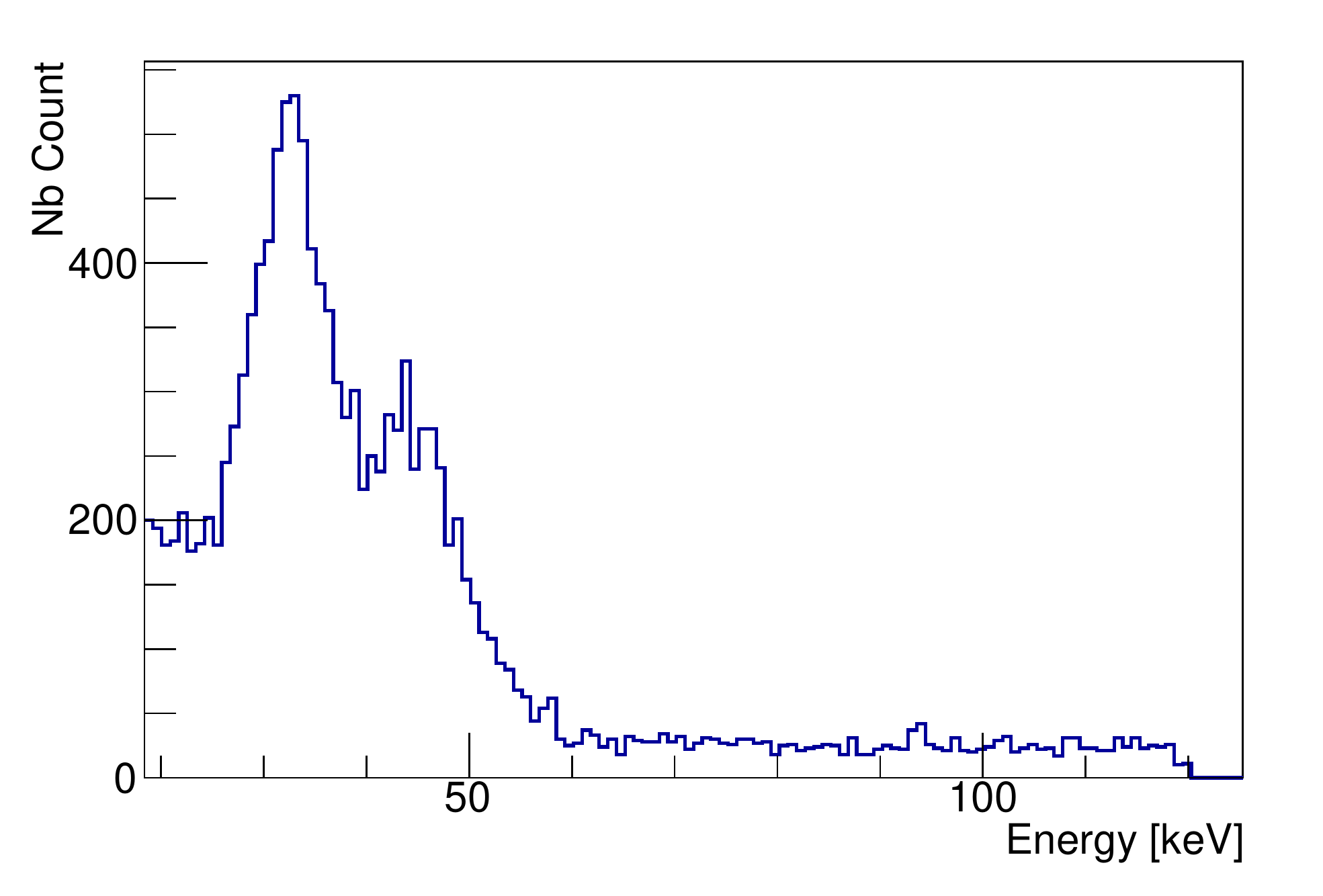}}
\end{minipage}
\begin{minipage}{0.5\linewidth}
\centerline{\includegraphics[width=\linewidth]{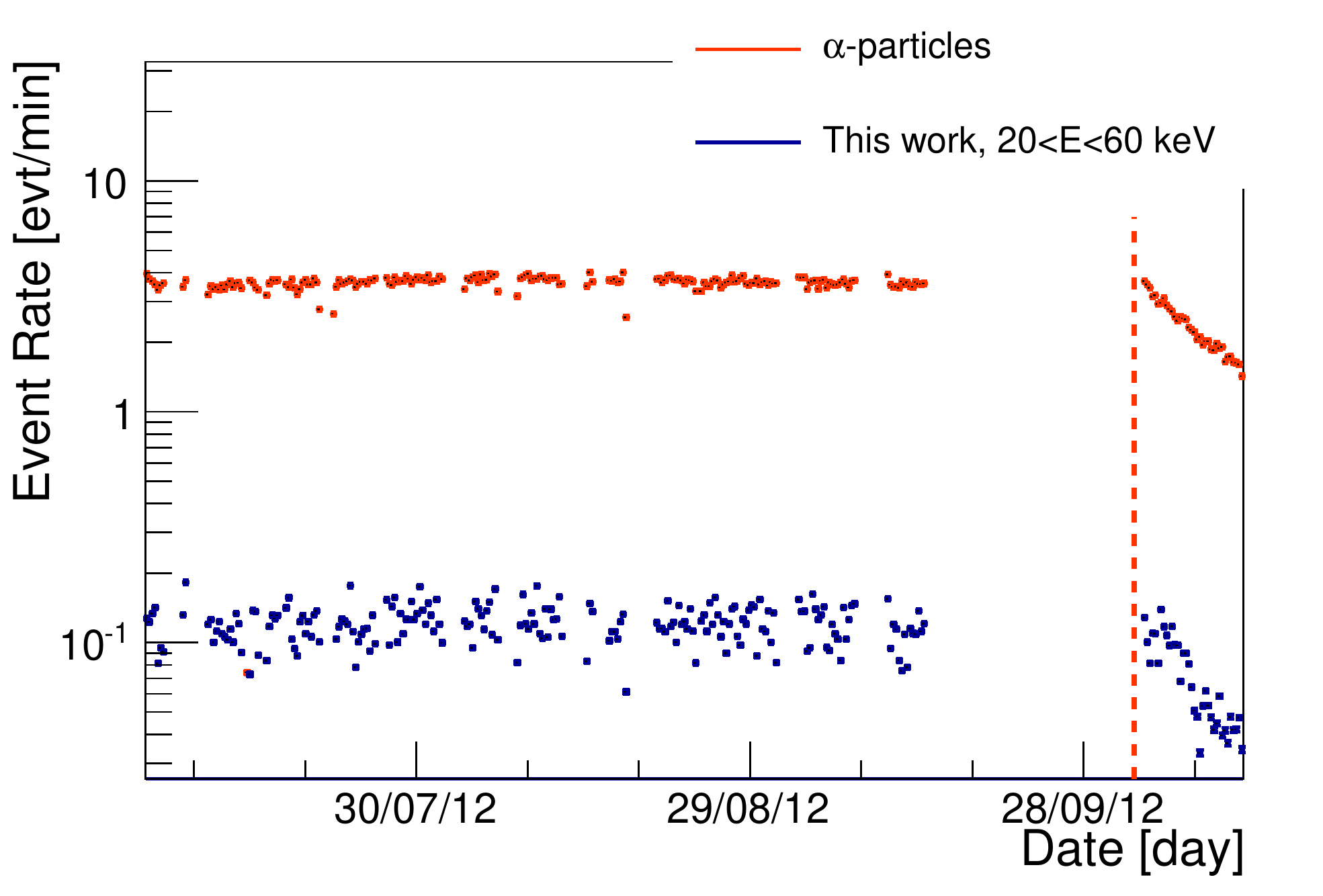}}
\end{minipage}
\caption{ Left:  The energy spectrum obtained after a preliminary analysis between 20 and 120 keV. Right:  The event rates of $\alpha$-particles (orange dots) and obtained with this analysis with $20<E<60$ keV (blue dots). The dashed orange line indicate the shutdown of the gas circulation system, on October $\mathrm{3^{rd}}$, 2012. }
\label{fig:ModaneInstal}
\end{figure}

The first available data set of the bi-chamber prototype was started on July $\mathrm{5^{th}}$ and finished on October $\mathrm{12^{th}}$ 2012. 
The measured total event rate of tracks with a 2 keV threshold was $5.6\pm 0.4\,\, \mathrm{evts/min}$ without any cuts. 
We performed a first data analysis 
that allowed us to obtain 3D-track events with an energy spectrum shown on Fig.~\ref{fig:ModaneInstal} left. In this spectrum, we can clearly identify two peaks at $32.4\pm 0.1$ and $44\pm 0.3$ keV, and a flat distribution above 60 keV.

Fig.~\ref{fig:ModaneInstal} right shows the event rate of $\alpha$-particles (Flash-ADC saturation above 120 keV) and of events with energies between 20 and 60 keV selected by this analysis.
These event rates remained constant until October $\mathrm{3^{rd}}$, when the gas circulation was stopped to isolate the bi-chamber from the gas circulation system. We can observe an exponential reduction of the event rates which fit with the $\mathrm{^{222}Rn}$ half-life $(T_{1/2} = 3.8\,\mathrm{days})$.   
It signs a pollution of the gas mixture by Radon isotopes contained in the circulation gas system originating from $\mathrm{^{238}U}$ and $\mathrm{^{232}Th}$ chains.
 The exponential decrease being dominated by the half-life of the  $\mathrm{^{222}Rn}$, hence, 
 the contribution from the decay of $\mathrm{^{220}Rn}$ may be neglected in a first order approximation.
The absence of $\mathrm{^{220}Rn}$ could be explained by its short half-life $(T_{1/2} = 55\,\mathrm{s})$ preventing it to reach the active volume. Indeed, the time required for $\mathrm{^{220}Rn}$ to reach the active volume of the chamber passing through the circulation circuit is at least one order of magnitude longer than its half-life.

As shown in Tab.~\ref{tab:Progeny}, the seven $\alpha$-particles from the decay of $\mathrm{^{222}Rn}$ and $\mathrm{^{220}Rn}$ descendants, called Radons progeny, are emitted with kinetic energies $E^{kin}_{\alpha}$ ranging from 5.5 to 8.8 MeV.
A simulation showed that these $\alpha$-particles can pass through the 24 \textmu m cathode to reach the other chamber even if the decay is produced in the gas volume. The simulated energy spectrum of the $\alpha$-particles reaching the other chamber was flat from 0 to 120~keV (our ionization energy dynamic range). 
In addition, if the decays occur at the cathode surface and if the $\alpha$-particles are absorbed in the matter, only the recoils of the daughter nuclei will be detected.
The recoils of the daughter nuclei from the Radons progeny are emitted, as shown on Tab.~\ref{tab:Progeny}, with kinetic energies from 100 to 170~keV.

There is a difference between the kinetic energy of an ion $E^{kin}_{recoil}$ and the energy released by ionization $E^{ioni}_{recoil}$. The Ionization Quenching Factor ({\it IQF}) is defined as the ratio between the ionization energy released by a recoil and the ionization energy released by an electron at the same kinetic energy.
Taking into account the {\it IQF} correction from SRIM~\cite{SRIM}, the recoils from $\mathrm{^{222}Rn}$ and $\mathrm{^{220}Rn}$ progeny would be detected with ionization energies from 38 to 70 keV.
It was shown that SRIM must overestimates the {\it IQF} roughly by 20 \% at such recoil energies~\cite{Guillaudin:2011hu}.
Thus, the measured ionization energies should be lowered by 20 \% with respect to the values on the Tab.~\ref{tab:Progeny}.

\begin{table}[h]
	\begin{center}   
		\begin{tabular}{|l|c|c|c|c|c|}
		\hline
		Parent & $E^{kin}_{\alpha}$ [MeV] & Daughter  & $E^{kin}_{recoil}$ [keV] & {\it IQF} [\%]& $E^{ioni}_{recoil}$ [keV] \\
      		\hline
     	 	\hline
     	 	\multicolumn{6}{|c|}{From $\mathrm{^{222}Rn}$} \\
     	 	\hline
     		 $\mathrm{^{222}Rn}$& $5.489$ & $\mathrm{^{218}Po}$ & $100.8$ & $37.93$ & $38.23$\\
      		$\mathrm{^{218}Po}$& $6.002$ & $\mathrm{^{214}Pb}$ & $112.3$  & $39.10$ & $43.90$\\
      		$\mathrm{^{214}Po}$& $7.687$ & $\mathrm{^{210}Pb}$ & $146.5$  & $40.12$ & $58.78$\\
      		\hline
		\hline
		 \multicolumn{6}{|c|}{From $\mathrm{^{220}Rn}$} \\
		\hline
      		$\mathrm{^{220}Rn}$& $6.288$ & $\mathrm{^{216}Po}$ & $116.5$ & $39.0$ & $45.4$\\
      		$\mathrm{^{216}Po}$& $6.778$ & $\mathrm{^{212}Pb}$ & $128.0$ & $39.1$ & $50.0$\\
	      	$\mathrm{^{212}Po}$& $8.785$ & $\mathrm{^{208}Pb}$ & $169.1$ & $41.3$ & $69.8$\\
		$\mathrm{^{212}Bi}$& $6.090$ & $\mathrm{^{208}Tl}$ & $117.2$ & $39.0$ & $45.7$\\
	      	\hline
		
		\end{tabular}
		\caption{\it This table details the $\alpha$-decays from $\mathrm{^{222}Rn}$ and $\mathrm{^{220}Rn}$ progeny, the kinetic energies $E^{kin}_{\alpha}$ of the emitted $\alpha$-particles and the kinetic  $E^{kin}_{recoil}$ and ionization $E^{ioni}_{recoil}$ energies of the emitted daughter nuclei. The {\it IQF} was estimated with SRIM.}
		\label{tab:Progeny}
	\end{center}
\end{table}

In conclusion, the 3D-track events associated to the energy spectrum shown on Fig.~\ref{fig:ModaneInstal} left come from the $\mathrm{^{222}Rn}$ progeny in the detector. 
Indeed, taking into account the {\it IQF} overestimation, the energy range of the recoils from the $\mathrm{^{222}Rn}$ progeny would fit the spectrum. 
The  identification of the contribution for each nuclear recoil from the $\mathrm{^{222}Rn}$ progeny will be studied in a future paper.

For directional detection it is important to discriminate Radon progeny events from WIMP like events. This discrimination should be possible by using the time and spatial coincidence between the two chambers. For these events, we expect one recoil event in one of the chambers and one recoil event in the other one.
For this data taking, the time synchronization of the chambers was not available.
The study of the coincidences between the two chambers will be performed with a new 12 \textmu m cathode for the next data taking period.

%%%%%%%%%%%%%%%%%%%%%%%%%%%%%%%%%%%%%%%%%%%%%%%%

\section{Conclusions and Perspectives}
\label{sec:Perspectives}

The MIMAC bi-chamber prototype was installed at the LSM in June 2012.
A preliminary analysis of the first data set allowed us to observe, for the first time, the 3D-track and energy of recoils from the $\mathrm{^{222}Rn}$ progeny. At present, only the $\alpha$ particle spectrum from $\mathrm{^{222}Rn}$ progeny was measured~\cite{Burgos:2007tt}.
This kind of measurement shows the potential of the MIMAC experiment for 3D recoil track measurements at low ionization energy.
Moreover, using the time correlation between the two chambers of the bi-chamber module, we should be able to discriminate these events.

The next step of the MIMAC project  will be the development of the MIMAC $1\,\mathrm{m^3}$. This detector will be the demonstrator for a large TPC devoted to DM directional search.


\section*{References}

\begin{thebibliography}{99}

\bibitem{Spergel:1987kx}
  D.N.~Spergel,
  \Journal\PRD{\bf 37}{1353}{1988}.

\bibitem{billard:2012ProfLikeli} J. Billard {\it et al.}, \Journal{\PRD}{85}{035006}{2012}.

\bibitem{Ahlen:2009ev}
  S.~Ahlen {\it et al.},
  Int.\ J.\ Mod.\ Phys.\ A {\bf 25}  1 (2010). 


\bibitem{Iguaz:2011yc}
  F.~J.~Iguaz {\it et al.}, JINST {\bf 6} P07002 (2011). 
  
\bibitem{Bourrion:2011vk}
  O.~Bourrion {\it et al.},
  EAS Publ.\ Ser.\  {\bf 53} 129  (2012).
    
\bibitem{Billard:2013cxa}
  J.~Billard {\it et al.},
    arXiv:1305.2360 (2013).
 
 \bibitem{Billard:2012bk}
  J.~Billard {\it et al.}, 
  JCAP {\bf 1204} 006 (2012).

\bibitem{Guillaudin:2011hu}
  O.~Guillaudin {\it et al.},
  EAS Publ.\ Ser.\  {\bf 53} 119 (2012).

  
 \bibitem{SRIM}
  J.F.~ Ziegler {\it et al.},
  Pergamon Press New York, www.srim.org (1985).
  
  

\bibitem{Burgos:2007tt}
  S.~Burgos {\it et al.},
  \Journal\NIMA{ 584} {114}{2008}.
  

\end{thebibliography}
\end{document}